\documentclass[12pt]{article}
\usepackage[usenames,dvipsnames]{color}
\usepackage{amsmath,amsthm,amsfonts,amssymb,multicol,amscd,amsbsy}
\usepackage{graphicx}
\usepackage{booktabs}
\usepackage{array}
\usepackage{enumerate}
\usepackage{url}
\usepackage[nodayofweek]{datetime}
\usepackage[english]{babel}
\usepackage{mathtools}
\usepackage[toc,page]{appendix}
\usepackage{verbatim} 
\usepackage{amsthm} 
\usepackage{enumitem}
\usepackage{enumerate}
\usepackage{bbm} 
\usepackage[normalem]{ulem} 
\usepackage{bm}
\usepackage{hyperref}
\usepackage{latexsym}
\usepackage[font=footnotesize,labelfont=bf]{caption}
\usepackage{authblk}

\newcommand{\be}{\begin}
\newcommand{\e}{\end}
\newcommand{\beq}{\begin{equation}}
\newcommand{\eeq}{\end{equation}}

\newcommand{\ul}{\underline}
\renewcommand{\l}{\left}
\renewcommand{\r}{\right}

\renewcommand{\Re}{\mathrm{Re}}

\newcommand{\set}[1]{\mathbb{#1}}
\newcommand{\curly}[1]{\mathcal{#1}}

\newcommand{\setof}[2]{\left\{ #1\; : \;#2 \right\}}

\newcommand{\C}{\set{C}}
\newcommand{\Z}{\set{Z}}

\newcommand{\lam}{\lambda}
\newcommand{\Lam}{\Lambda}
\newcommand{\gam}{\gamma}

\newcommand{\al}{\alpha}
\newcommand{\de}{\delta}

\newcommand{\scp}[2]{\langle#1,#2\rangle}

\newcommand{\ket}[1]{|#1\rangle}

\newcommand{\ketbra}[2]{ |#1 \rangle \langle #2|}

\theoremstyle{definition}

\numberwithin{equation}{section}

\theoremstyle{remark}

\def\dotuline{\bgroup
  \ifdim\ULdepth=\maxdimen  
   \settodepth\ULdepth{(j}\advance\ULdepth.4pt\fi
  \markoverwith{\begingroup
  \advance\ULdepth0.08ex
  \lower\ULdepth\hbox{\kern.15em .\kern.1em}%
  \endgroup}\ULon}

\def\dashuline{\bgroup
  \ifdim\ULdepth=\maxdimen  
   \settodepth\ULdepth{(j}\advance\ULdepth.4pt\fi
  \markoverwith{\kern.15em
  \vtop{\kern\ULdepth \hrule width .3em}%
  \kern.15em}\ULon}
\allowdisplaybreaks

\usepackage{tikz}

\usetikzlibrary{decorations.pathreplacing,decorations.markings}
\usepackage{subfig}
\usepackage{xcolor}
\usepackage{xspace}

\tikzset{
  mid arrow/.style={postaction={decorate,decoration={
        markings,
        mark=at position .5 with {\arrow[#1]{stealth}}
      }}},
}

\newcommand{\hlr}{\tikz{\draw[thick] (0,-.2)--(0.2,-.2)--(0.2,0);\draw[fill=black] (0,-.2) circle (1pt);\draw[fill=black] (0.2,-0.2) circle (1pt);\draw[fill=black] (0.2,0) circle (1pt);}}

\newcommand{\hrr}{\tikz{\draw[thick] (0,0)--(0.2,0)--(0.4,0);\draw[fill=black] (0,0) circle (1pt);\draw[fill=black] (0.2,0) circle (1pt);\draw[fill=black] (0.4,0) circle (1pt);}}

\newcommand{\oll}{\tikz{\draw[thick,->] (0,0)--(0,0.2); \draw[thick,->] (0,0)--(0.2,0);\draw[fill=black] (0,0) circle (1pt);\draw[fill=black] (0.2,0) circle (1pt);\draw[fill=black] (0,0.2) circle (1pt);}}

\newcommand{\oul}{\tikz{\draw[thick,->] (0,-.2)--(0,0); \draw[thick,->](0,0)--(0.2,0);\draw[fill=black] (0,0) circle (1pt);\draw[fill=black] (0.2,0) circle (1pt);\draw[fill=black] (0,-0.2) circle (1pt);}}

\newcommand{\olr}{\tikz{\draw[thick,->] (0,-.2)--(0.2,-.2); \draw[thick,->] (0.2,-0.2)--(0.2,0);\draw[fill=black] (0,-.2) circle (1pt);\draw[fill=black] (0.2,-0.2) circle (1pt);\draw[fill=black] (0.2,0) circle (1pt);}}

\newcommand{\our}{\tikz{\draw[thick,->] (0,0)--(0.2,0); \draw[thick,->](0.2,-.2)--(0.2,0);\draw[fill=black] (0,0) circle (1pt);\draw[fill=black] (0.2,0) circle (1pt);\draw[fill=black] (0.2,-0.2) circle (1pt);}}

\begin{document}
\title{Gapped PVBS models for all species numbers and dimensions}
\date{June 3, 2019}

\author[1]{Marius Lemm}
\author[2]{Bruno Nachtergaele}

\affil[1]{\textit{Department of Mathematics\\
Harvard University\\
Cambridge, MA 02138, USA}}

\affil[2]{\textit{Department of Mathematics and Center for Quantum Mathematics and Physics\\
University of California, Davis\\
Davis, CA 95616, USA}}


\maketitle

\begin{abstract}
  Product vacua with boundary states (PVBS) are cousins of the Heisenberg XXZ spin model and feature $n$ particle species on $\Z^d$. The PVBS models were originally introduced as toy models for the classification of ground state phases. A crucial ingredient for this classification is the existence of a spectral gap above the ground state sector. In this work, we derive a spectral gap for PVBS models at arbitrary species number $n$ and in arbitrary dimension $d$ in the perturbative regime of small anisotropy parameters. Instead of using the more common martingale method, the proof verifies a finite-size criterion in the spirit of Knabe. 
  \end{abstract}

\section{Introduction}
This paper concerns a family of frustration-free quantum spin models called ``Product Vacua with Boundary States'' (PVBS). These models were first introduced in \cite{BN} in one dimension and subsequently generalized to $d$ dimensions in \cite{BHNY}. The PVBS models can be seen as variants of the Heisenberg XXZ models in an external magnetic field. A remarkable feature of these is that, for appropriate parameter values, their gapped ground state phases can be explicitly classified in terms of edge-localized particles \cite{BHNY,BN,BNY}.

 Our goal in this paper is to prove that the PVBS models exhibit a spectral gap above the ground state for arbitrary species number and arbitrary dimension, in a perturbative regime with small values of the anisotropy parameters, which also decrease from species to species. We review previous results below, which are restricted to species numbers $n\leq 2$ in dimensions $d\geq 2$ \cite{B,BNY}. Instead of using the more common martingale method \cite{N}, our proof is based on finite-size criteria {\`a} la Knabe \cite{GM,K,L1,LM}, whose original criterion was inspired by the seminal proof that the AKLT chain is gapped \cite{AKLT}. Specifically, we build on a recent generalization of these criteria to arbitrary dimensions $d$ \cite{L1}. This is the first time that such finite-size criteria could be verified in a concrete model for dimensions $d>1$. The advantage of the more coarse method of finite-size criteria is that it handles the increase of complexity coming from large $n$ and $d$ seamlessly. A disadvantage is that, by design, the method requires a sufficiently large finite-size gap which currently limits us to the regime of small anisotropy parameters (see also \cite{L2} where the same phenomenon occurs in a one-dimensional context). The method does quantify what ``small'' means here; see \eqref{eq:cdn} below.
  
We focus on the case of periodic boundary conditions for simplicity, and we interpret the result as a bulk gap. We expect that the methods developed here can be extended to derive a spectral gap for systems with free boundary conditions along sufficiently nice boundary shapes, like boxes; see Remark \ref{rmk:bc}. We emphasize that the issue of boundary conditions is subtle for PVBS models, since they are known to exhibit weakly excited edge modes for certain half-plane geometries \cite{BNY}.

For motivation and background concerning spectral gaps, we refer to \cite{LM,NSY}. Here we just mention two facts: (1) Spectral gaps are fundamental for distinguishing quantum phases of matter via Hastings' spectral flow (also called quasi-adiabatic evolution) \cite{H}, see also \cite{BMNS,NSY}. (2) The stability of spectral gaps under various ``local'' perturbations of the Hamiltonian has been a topic of interest recently \cite{BHM,FP,MZ,MN}, and it is of course beneficial to have a wide net of gapped Hamiltonians available for further stability analysis.

\subsection{Setup}
The PVBS models are defined in terms of various parameters as follows. As the underlying graph, we take $\Z^d$, respectively a box $\Lam_L=((-L,L]\cap\Z)^d
$ with periodic boundary conditions. The PVBS model with species number $n\geq 1$ is then defined on the Hilbert space
$$
\curly{H}_L:=\bigotimes_{x\in \Lam_L} \C^{n+1}.
$$
Its local interactions are defined in terms of a collection of $n\times d$ positive anisotropy parameters (one for every species and every direction), which we denote by $\{\lam_{i}^{(k)}\}_{\substack{1\leq i\leq n\\ 1\leq k\leq d}}$. For a direction $1\leq k\leq d$, the local interaction term is given by 
\beq\label{eq:hkdefn}
h^{(k)}:=\sum_{i=1}^n \ketbra{\hat \phi_i^{(k)}}{\hat\phi_i^{(k)}}
+\sum_{1\leq i\leq j\leq n} \ketbra{\hat \phi_{ij}^{(k)}}{\hat\phi_{ij}^{(k)}}
\eeq
where the vectors $\hat \phi_i^{(k)},\hat \phi_{ij}^{(k)}\in \C^{n+1}\otimes \C^{n+1}$ are the normalized versions of
$$
\phi_i^{(k)}:=\ket{0\otimes i}-\lam_i^{(k)}\ket{i\otimes 0},
\qquad \phi_{ij}^{(k)}:=\lam_i^{(k)}\ket{i\otimes j}-\lam_j^{(k)}\ket{j\otimes i},
\qquad \phi_{ii}^{(k)}:=\ket{i\otimes i},
$$
with $1\leq i<j\leq n$ and $1\leq k\leq d$. (We remark that $h^{(k)}$ is itself a projection.) The full Hamiltonian $H_L$ is defined by placing an interaction $h^{(k)}$ at every oriented edge in direction $k$ of the periodic box $\Lam_L$, with the convention that \emph{edges are oriented in the direction of the canonical basis vectors} $e_1,\ldots,e_d$. The Hamiltonian thus reads
$$
H_L=\sum_{x\in \Lam_L} \sum_{k=1}^d  h^{(k)}_{x,x+e_k}.
$$
(We recall that we are using periodic boundary conditions.) The Hamiltonian $H_L$ is frustration-free, which can be phrased as 
$$
H_L\geq 0,\qquad \ker H_L=\bigcap_{x\in\Lam_L}\bigcap_{k=1}^d  \ker h^{(k)}_{x,x+e_k}\neq \{0\}.
$$
Indeed, notice that the tensor power $\ket{0}^{\otimes \Lam_L}$ is a ground state. For connected sets with free boundary conditions, the ground states are labeled by subsets $M\subset \{1,\ldots,N\}$ and are given in \cite{BHNY}. Our main object of interest is the \emph{spectral gap} $\gam_L$ of $H_L$, which is defined as the smallest positive eigenvalue of $H_L$. 

In summary, the frustration-free PVBS models are defined in terms of the following parameters:

\be{itemize}
\item dimension $d\geq 1$,
\item species number $n\geq 1$,
\item $nd$ positive anisotropy parameters $\{\lam_{i}^{(k)}\}_{\substack{1\leq i\leq n\\ 1\leq k\leq d}}$.
\e{itemize}

\subsection{Main result}
A central question concerning the PVBS models is whether they exhibit a spectral gap above the ground state, which we will define to mean $\inf_L \gam_L>0$. This is expected to hold at least as long as there is sufficient anisotropy (which translates to $\lam_i^{(k)}\neq 1$ for $1\leq i\leq n$ and $1\leq k\leq d$). Under this assumption, a spectral gap has been established for the following parameter regimes

\be{itemize}
\item For $d=1$ and $n\geq 1$ arbitrary \cite{BN}. 
\item For $d\geq 2$ and $n=1$ \cite{BHNY} and, more recently, $n=2$ \cite{B}.
\e{itemize}

In all these works, the spectral gap is derived via the martingale method \cite{N}. From the works \cite{BHNY,B}, it appears that the martingale method becomes rather cumbersome in dimensions $d\geq 2$, especially for higher species numbers $n$. Instead, we use finite-size criteria which are more flexible, but also more coarse, and so the result only holds for small values of the anisotropy parameters in the following sense.

We shall write $0<\de<1$ for the small parameter. We assume that
\beq\label{eq:assumedelta}
\frac{1}{2} \de^i\leq \lam_{i}^{(k)} \leq 2 \de^i,
\eeq
holds for all $1\leq i\leq n$ and $1\leq k\leq d$. 

Our main result establishes a spectral gap for PVBS models in dimensions $d\geq 2$ at arbitrary species number $n$ for sufficiently small values of $\de>0$.

\be{thm}[Spectral gap]
\label{thm:main}
Fix two integers $d\geq 2$ and $n\geq 1$. There exists a positive constant $c_{d,n}>0$ such that for all $0<\de<c_{d,n}$, the following holds. If the anisotropy parameters $\{\lam_{i}^{(k)}\}_{\substack{1\leq i\leq n\\ 1\leq k\leq d}}$ satisfy \eqref{eq:assumedelta}, then $H_L$ is gapped, i.e.,
\beq
\inf_{L\geq 2}\gam_L>0.
\eeq
\e{thm}

\be{rmk}
\be{enumerate}[label=(\roman*)]
\item 
The proof of Theorem \ref{thm:main} yields explicit numerical choices for the constant $c_{d,n}$. For instance, when $d\geq 3$, one can take 
\beq\label{eq:cdn} 
c_{d,n}=\frac{1-\frac{3}{m}}{96n^2(m+1)^d}>0,
\eeq
with $m=\max\{n^{1/(d-2)},3d\}$. The choice \eqref{eq:cdn} is far from optimal in terms of the universal constants, but it possesses the optimal scaling in $m$ and $n$ afforded by the proof. See \eqref{eq:2dchoice} for a possible choice of $c_{2,n}$.
\item There is some flexibility regarding assumption \eqref{eq:assumedelta}. First, the constant $2$ on both sides can be changed to a number $a>0$ and the only effect of this is to rescale $c_{d,n}$ by $\l(\frac{2}{a}\r)^{2}$. Second, the main role of \eqref{eq:assumedelta} is to ensure that, for $i<j$, we have $\lam_{j}^{(k)}/\lam_{i}^{(k)}\to 0$ as $\de\to 0$. Replacing \eqref{eq:assumedelta} by this weaker assumption also yields Theorem \ref{thm:main}, but without any quantitative control on $c_{d,n}$.
\e{enumerate}
\e{rmk}

\subsection{Comments on the proof strategy}
\label{sect:strategy}
Our proof is based on verifying a finite-size criterion for spectral gaps in the spirit of \cite{GM,K,L1,LM}. In general a finite-size criterion says that if the Hamiltonian has a sufficiently large gap on a certain finite subsystem, then it is gapped uniformly for all system sizes. The choice of the appropriate subsystems becomes a delicate matter in higher dimensions, and we build on a recent observation \cite{L1} that there is additional freedom if one applies the operator Cauchy-Schwarz inequality in the appropriate place.

 In Section \ref{sect:fs1}, we present the particular finite-size criterion we use (Theorem \ref{thm:fs}), which is tailor-made for the application to the PVBS model. Let us explain what we mean by this. A key feature of our assumption \eqref{eq:assumedelta} is that the Hamiltonian with $\de=0$ is a sum of commuting projections and therefore has spectral gap $=1$. Our result is then based on the idea that taking $\delta>0$ amounts to a perturbation of norm $\leq C\delta$, which then does not close the gap (and hence verifies the finite-size criterion) for $\de$ sufficiently small. 
 
 However, and this is the main technical problem we face, to implement this heuristic idea rigorously, it is necessary to ensure that there are no ground state degeneracies present in the finite system which are lifted by introducing $\de$ and would thus lead to small excitation energies of $O(\de)$, thereby spoiling the $O(1)$ gap. Here is a simple example of this phenomenon (which is a common problem faced in degenerate perturbation theory) occurring for a pair of ``frustration-free'' $3\times 3$ matrices. Set
$$
A=\left(\be{array}{ccc} 1&0&0\\	0&0&0\\ 0&0&0 \e{array}	\right),\qquad 
B=\left(\be{array}{ccc} 1&0&0\\	0&\de&0\\ 0&0&0 \e{array}	\right).
$$
Then $A$ has a spectral gap of $1$, and $\|B-A\|=\de$, but the spectral gap of $B$ is equal to $\de$, i.e., small. The key fact about the finite subsystem that we use for the finite-size criterion (which we call ``box on a stick''; see Section \ref{sect:fs} below) is that it \emph{avoids any artificial ground state degeneracies}. More precisely, all the degeneracies that occur for the box on a stick can be labeled using particle number symmetry and hence we can apply non-degenerate perturbation theory in each particle number sector. This allows us to verify the finite-size criterion for the box on a stick by the perturbative argument in the small parameter $\de$ that was sketched above; see Section \ref{sect:verify}. (For contrast, we mention that a simple box without stick does have a large number of artificial degeneracies in each particle number sector, so the perturbative argument breaks down in that case.) 

We close the introduction with a remark about a possible generalization of Theorem \ref{thm:main}.

\be{rmk}[The case of free boundary conditions]
\label{rmk:bc}
We expect that the argument can be generalized to derive a spectral gap with free boundary conditions on $\Lam_L$, instead of periodic ones. We mention that the key idea for this generalization is that, near the part of the boundary where one cannot fit a box on a stick into the set without losing the stick, one replaces the box by a cuboid, such that the stick still fits into $\Lam_L$. This may require rotating the stick by $\pi/2$. This change affects the combinatorics used to establish Theorem \ref{thm:fs} near the boundary, but we fully expect that one still obtains a gap threshold of order $\frac{1}{m}$. Finally, the resulting finite-size criterion can be verified for the ``cuboid on the stick`` that one gets near the boundary, since, as we will see in proving Proposition \ref{prop:verify}, the existence of the stick is what is crucial to obtain a large gap of the subsystem, for sufficiently small $\de>0$.
\e{rmk}

\section{A finite-size criterion}\label{sect:fs1}
The main result of this section is the finite-size criterion Theorem \ref{thm:fs} which holds for arbitrary frustration-free Hamiltonians defined on a periodic box $\Lam_L\subset \Z^d$ (with directed or undirected edges). 

\subsection{Subsystem Hamiltonians}
\label{sect:fs}
Given a connected subgraph $\curly{G}\subset \Z^d$, let $\curly{E}_{\curly{G}}$ be the set of edges in $\curly{G}$. We consider these sets as oriented subgraphs of $\Z^d$ with the same orientation as before (i.e., edges are oriented along the canonical basis vectors).  We can then define the subsystem Hamiltonian associated to $\curly{G}$ concisely as 
\beq\label{eq:HSdefn}
H_\curly{G}:=\sum_{e\in \curly{E}_{\curly{G}}} h_e.
\eeq
 Here and in the following, we will often suppress the superscript $(k)$ when denoting the interaction $h^{(k)}$, as it will be clear from the direction of the edge under consideration. I.e., we identify
\beq\label{eq:hkconvention}
h_e=h^{(k)}_{e}, \qquad \textnormal{if $e$ is parallel to $e_k$}.
\eeq 

From now on, we fix a dimension $d\geq 2$ and a species number $n\geq 1$. The subgraphs we use for the finite-size criterion are the following $\curly{C}_m$, which we call ``box on a stick''. We set $\curly{C}_m=\curly{S}\cup \curly{B}_m$ where $\curly{S}$ is the stick
$$
\curly{S}:=\{0,e_1,\ldots,(n-1)e_1\}
$$
and $\curly{B}_m$ is the box of sidelength $m$ defined by
$$
\curly{B}_m:=\setof{\sum_{k=1}^d a_k e_k}{n\leq a_1\leq m+n,\textnormal{ and }0\leq a_k\leq m,\,\forall k\geq 2}.
$$

To these subsystems, we associate a frustration-free subsystem Hamiltonian $H_{\curly{C}_m}$ via Definition \eqref{eq:HSdefn}, and we write $\gam_{\curly{C}_m}$, for its spectral gap. 

\subsection{The finite-size criterion}
We are now ready to state the finite-size criterion which lower bounds the spectral gap of $H_L$ in terms of the spectral gap of the box on a stick, $H_{\curly{C}_m}$. 

\be{thm}[Finite-size criterion]\label{thm:fs}
Let $m\geq 4$ and $L\geq 2m+1$. Then
\beq\label{eq:fs}
\gam_L\geq  \frac{m^d}{(m+1)^{d}+n}
\l(\gam_{\curly{C}_m}-\frac{1}{m}-\frac{8d}{m^2}\l(1+\frac{n}{m^{d-2}}\r)\r).
\eeq
\e{thm}

\be{rmk}
\be{enumerate}[label=(\roman*)]
\item This theorem functions as a finite-size criterion in the following way: If $\gam_{\curly{C}_m}>\frac{1}{m}+\frac{3d}{m^2}\l(1+\frac{n}{m^{d-2}}\r)$ for some $m\geq 4$, then $\inf_L\gam_L>0$, and so $H_L$ is gapped. We will verify this condition later for $\de$ sufficiently small. 

\item The proof of Theorem \ref{thm:fs} closely follows the proof of a similar criterion just for boxes $\curly{B}_m\subset \Z^d$ from \cite{L1}. The key idea introduced there, which allows to extend the previous results for $d\in \{1,2\}$ to arbitrary dimensions, is to employ the operator Cauchy-Schwarz inequality \eqref{eq:cs} to allow for more flexibility in Knabe's original combinatorial argument \cite{K}. Here, we follow that approach but we add the stick to the box and observe that this does not change the gap threshold significantly as long as $n\ll m^{d-1}$, as one can see from \eqref{eq:fs}. Adding the stick is crucial for applying the finite-size criterion to the PVBS model at hand, since, as anticipated in the introduction, the box on a stick has no accidental ground state degeneracies for the unperturbed system with $\de=0$.

\item While we formulate Theorem \ref{thm:fs} for the PVBS model for simplicity, the statement and proof extend verbatim to any frustration-free model defined on a periodic box $\Lam_L\subset \Z^d$.

\e{enumerate}
\e{rmk}

In the remainder of this section, we prove Theorem \ref{thm:fs}

\subsection{Step 1: Squaring the Hamiltonian}
We introduce some convenient notation. We write $\curly{E}_L=\curly{E}_{\Lam_L}$ for the set of oriented edges in the box $\Lam_L$ with periodic boundary conditions. Given two edges $e,e'\in\curly{E}_L$, we write $e\sim e'$ if $e$ and $e'$ share exactly one vertex, and we write $e\not\sim e'$ if they share no vertex. 
Finally, we denote the anticommutator of two operators $A,B$ by $\{A,B\}=AB+BA$.

The first step to prove Theorem \ref{thm:fs} is to square the Hamiltonian $H_L$. Using that the local interactions $h^{(k)}$ are projectors (so $h_e^2=h_e$), we obtain
\beq\label{eq:Hsquare}
H_L^2=H_L+Q+R,
\eeq
where we introduced
\beq\label{eq:QRdefn}
Q:=\sum_{\substack{e,e'\in \curly{E}_L\\ e\sim e'}}   \{h_{e},h_{e'}\},
\qquad 
R:=\sum_{\substack{e,e'\in \curly{E}_L\\ e\not\sim e'}}  \{h_{e},h_{e'}\}.
\eeq

\subsection{Step 2: Comparing with an auxiliary operator}
We define the shifted system
$$
x+\curly{C}_m:=\setof{y\in \Lam_L}{y-x\in\curly{C}_m},
$$
where $y-x$ is defined with periodic boundary conditions on $\Lam_L$. Now we introduce the auxiliary operator
$$
A:=\sum_{x\in \Lam_L} H_{x+\curly{C}_m}^2
$$ 
where we distributed the subsystem Hamiltonians across the entire box $\Lam_L$ via translation, always incorporating periodic boundary conditions for $\Lam_L$. 

\be{prop}\label{prop:knabe}
It holds that
\begin{align}
\label{eq:A1}
A\geq&\, \gam_{\curly{C}_m}m(m+1)^{d-1}H_L,\\
\label{eq:A2}
A\leq& \l(m(m+1)^{d-1}+n+ 8d\l((m+1)^{d-2}+n\r)\r)H_L\\
\nonumber
&+ \l(m^2 (m+1)^{d-2}+n\r) (Q+R).
\end{align}
\e{prop}

We remark that the relevant regime to keep in mind is that of large $m$.

\be{proof}[Proof of Proposition \ref{prop:knabe}]
Fix $x\in \Lam_L$. We may compute, as in \eqref{eq:Hsquare} above,
$$
H_{x+\curly{C}_m}^2=H_{x+\curly{C}_m}+Q_{x+\curly{C}_m}+R_{x+\curly{C}_m}. 
$$
Here we defined $Q_{x+\curly{C}_m},R_{x+\curly{C}_m}$ in the same way as $Q,R$ in \eqref{eq:QRdefn} above, except that the edges $e,e'$ are restricted to $x+\curly{C}_m$, viewed as an oriented subgraph of $\Z^d$. We may reorganize the terms comprising the operator $A$ as follows: We group together the following types of terms: (a) terms $h_e$, (b) terms $\{h_e,h_{e'}\}$ with $e\sim e'$ and (c) terms $\{h_e,h_{e'}\}$ with $e\not\sim e'$.  The result is
$$
\begin{aligned}
A=\sum_{x\in \Lam_L} 
H_{x+\curly{C}_m}
+\sum_{x\in \Lam_L} 
Q_{x+\curly{C}_m}
\sum_{x\in \Lam_L} 
R_{x+\curly{C}_m}
=:(a)+(b)+(c).
\end{aligned}
$$
We now count how often each individual term of types (a)-(c) appears in the above sums.\\

\dashuline{Type (a) terms.} Modulo translations and rotations, there are only two different types of (a) terms: those labeled with edges in the $e_1$ direction, we call this the \emph{vertical} direction, and those with edges labeled in the remaining directions $e_2,\ldots,e_d$; we call these \emph{non-vertical terms}. We first count the non-vertical, i.e., we count the number of translated boxes $\curly{B}_m$ which contain a fixed edge, $e_2$, say. This count is $m(m+1)^{d-1}$. A vertical edge appears, additionally, in $n-1$ sticks. Since each term $h_e\geq 0$, these combinatorial considerations prove the operator inequality
\beq\label{eq:Hest}
m(m+1)^{d-1}H_L\leq \sum_{x\in \Lam_L} H_{x+\curly{C}_m}\leq \l(m(m+1)^{d-1}+n\r)H_L.
\eeq
The lower bound in \eqref{eq:Hest} already implies \eqref{eq:A1}. Indeed, by frustration-freeness, the spectral theorem, and translation invariance, we have $H_{x+\curly{C}_m}^2\geq \gam_{\curly{C}_m} H_{x+\curly{C}_m}$ and so
$$
A\geq  \gam_{\curly{C}_m}\sum_{x\in \Lam_L} H_{x+\curly{C}_m}\geq \gam_{\curly{C}_m}m(m+1)^{d-1}H_L,
$$
which is \eqref{eq:A1}.

We continue with the proof of \eqref{eq:A2}; for this we also need to count the type (b) and (c) terms.\\

\dashuline{Type (b) terms.}
These are the crucial ones because they are not necessarily positive definite. We summarize the result of the computation in a lemma.
\be{lm}\label{lm:Q}
We have the operator inequaliy
\beq\label{eq:Qest}
\sum_{x\in \Lam_L} Q_{x+\curly{C}_m}\leq \l(m^2 (m+1)^{d-2}+n\r) Q + 8d\l((m+1)^{d-2}+n\r)H_L.
\eeq
\e{lm}

\be{proof}[Proof of Lemma \ref{lm:Q}]
 There are various basic kinds of type (b) terms $\{h_e,h_{e'}\}$. We recall that the $e_1$-direction is the vertical direction for us. The first distinction we make is whether whether either of the edges $e,e'$ is vertical, which leads to the decomposition
$$
Q=Q_{0v}+Q_{1v}+Q_{2v},
$$
where $Q_{iv}$ contains the (b) terms $\{h_e,h_{e'}\}$ where exactly $i\in \{0,1,2\}$ of the edges $e\sim e'$ are vertical (and we use the same convention to define the terms $Q_{x+\curly{C}_m,iv}$ for $i\in \{0,1,2\}$).\\

\textbf{$Q_{0v}$ terms.} We first count the occurrences of each term from $Q_{0v}$ in the sum $\sum_{x\in \Lam_L} Q_{x+\curly{C}_m,0v}$. For this we can restrict attention to the box $\curly{B}_m$. Modulo translations and rotations, there are two basic kinds of such terms, depending on the relative positions of $e,e'$, and we denote these by $\hrr$ and $\hlr$, respectively. We count that there are $(m-1)(m+1)^{d-1}=(m^2-1) (m+1)^{d-2}$ $\hrr$ terms and $m^2(m+1)^{d-2}$  $\hlr$ terms. A crucial observation, which we borrow from \cite{L1}, is that these counts only differ by $(m+1)^{d-2}$, a lower order term. The difference is controlled using the operator Cauchy-Schwarz inequality in the form
\beq\label{eq:cs}
-\{h_e,h_{e'}\}=-h_eh_{e'}-h_{e'}h_e\leq (-h_e)^2+h_{e'}^2=h_e+h_{e'},
\eeq
which implies the bound
$$
-Q_{0v}^{\hrr}\leq 2\sum_{e \textnormal{ non-vert.}} h_e\leq 2H_L.
$$
We conclude
$$
\sum_{x\in \Lam_L} Q_{x+\curly{C}_m,0v} \leq m^2(m+1)^{d-2} Q_{0v}+2(m+1)^{d-2}H_L
$$

\textbf{$Q_{1v}$ terms.} Next, we count the occurrences of $Q_{1v}$ terms (i.e., $\{h_e,h_{e'}\}$ terms with exactly one vertical edge $e$ or $e'$) in $\sum_{x\in \Lam_L} Q_{x+\curly{C}_m,1v}$. Now we need to also account for rotational degrees of freedom, and so we consider $4$ subtypes of edge pairs, modulo rotations along the vertical axis: $\our,\olr,\oll$ and $\oul$, where the vertical edge is drawn vertically. Notice that the count for subtypes $\our,\olr,\oll$ is equal to the count we obtained for $Q_{0v}$, since all these edge pairs can only occur in the box $\curly{B}_m$ anyway. However, the $\oul$ subtype appears $1$ additional time compared to $Q_{0v}$, at the connection of the stick $\curly{S}$ with the box $\curly{B}_m$. We bound the missing terms via \eqref{eq:cs} which yields the operator inequality
$$
-Q_{1v}^{\our}-Q_{1v}^{\olr}-Q_{1v}^{\oll}\leq  3(d-1)H_L.
$$
(We mention that there is some overcounting in this bound, since only vertical edges appear $3(d-1)$ times after applying \eqref{eq:cs} to the left-hand side.) We conclude
$$
\sum_{x\in \Lam_L} Q_{x+\curly{C}_m,1v} \leq \l(m^2(m+1)^{d-2}+1\r) Q_{1v}+3(d-1)(m+1)^{d-2}H_L.
$$

\textbf{$Q_{2v}$ terms.} The count for these terms is equal to the count of $\hrr$ terms in $Q_0$, which was $(m^2-1) (m+1)^{d-2}$, plus the contribution from the stick, which is $n$. (Notice that this includes a pair of adjacent edges $e,e'$ where $e\in \curly{S}$ and $e'\in \curly{B}_m$.) We conclude that
$$
\sum_{x\in \Lam_L} Q_{x+\curly{C}_m,2v}= \l((m^2-1) (m+1)^{d-2}+n\r) Q_{2v}
$$

In order to obtain $Q=Q_{0v}+Q_{1v}+Q_{2v}$, we apply \eqref{eq:cs} again to the differences between $Q_{0v},Q_{1v},Q_{2v}$ and conclude
\beq
\sum_{x\in \Lam_L} Q_{x+\curly{C}_m}\leq \l(m^2 (m+1)^{d-2}+n\r) Q + 8d\l((m+1)^{d-2}+n\r)H_L.
\eeq
This proves Lemma \ref{lm:Q}.
\e{proof}

We continue with the proof of \eqref{eq:A2} in Proposition \ref{prop:knabe}; it remains to compute the type (c) terms.\\

\dashuline{Type (c) terms.}
First, observe that each type (c) term is non-negative since $\{h_e,h_{e'}\}\geq 0$ if $e\not\sim e'$, because in that case $h_e$ and $h_{e'}$ are commuting projectors. It then suffices to observe that the count of all type (c) terms is controlled by the largest count of type (b) terms, which as we saw above was $\max\{(m^2-1) (m+1)^{d-2}+n,m^2(m+1)^{d-2}+1\}$. Thanks to $\{h_e,h_{e'}\}\geq 0$, we can include the missing terms to find
\beq\label{eq:Rest}
\sum_{x\in \Lam_L} R_{x+\curly{C}_m}\leq \l(m^2 (m+1)^{d-2}+n\r) R.
\eeq
We recall that
$$
A=\sum_{x\in \Lam_L} H_{x+\curly{C}_m}^2
=\sum_{x\in \Lam_L} H_{x+\curly{C}_m}+\sum_{x\in \Lam_L} Q_{x+\curly{C}_m}+\sum_{x\in \Lam_L} R_{x+\curly{C}_m},
$$
and so the claim \eqref{eq:A2} follows by combining \eqref{eq:Hest}, \eqref{eq:Qest}, and \eqref{eq:Rest}.
\e{proof}

\subsection{Conclusion}
We are now ready to prove the finite-size criterion Theorem \ref{thm:fs}. For convenience, we abbreviate
$$
\al:=n+8d((m+1)^{d-2}+n).
$$
We combine \eqref{eq:Hsquare} and Proposition \ref{prop:knabe} to find
$$
\begin{aligned}
H_L^2
=&H_L+Q+R\\
\geq& H_L+\frac{A-\l(m(m+1)^{d-1}+\al\r)H_L}{m^2(m+1)^{d-2}+n} \\
\geq& H_L+\frac{m(m+1)^{d-1}(\gam_{\curly{C}_m}-1)-\al}{m^2(m+1)^{d-2}+n} H_L,\\
=& \frac{m(m+1)^{d-1}H_L}{m^2(m+1)^{d-2}+n}\\
& \times \l(\gam_{\curly{C}_m}-1+\frac{m^2(m+1)^{d-2}+n}{m(m+1)^{d-1}}-\frac{\al}{m(m+1)^{d-1}}\r).
\end{aligned}
$$
We compute the main contribution to the gap threshold:
$$
-1+\frac{m^2(m+1)^{d-2}}{m(m+1)^{d-1}}=-\frac{1}{m+1}\geq -\frac{1}{m},
$$
and the correction term
$$
\frac{n-\al}{m(m+1)^{d-1}}
=-\frac{8d((m+1)^{d-2}+n)}{m(m+1)^{d-1}}
\geq \frac{-8d}{m^2}\l(1+\frac{n}{m^{d-2}}\r).
$$
This proves
$$
\begin{aligned}
H_L^2
\geq& \frac{m(m+1)^{d-1}}{m^2(m+1)^{d-2}+n}
\l(\gam_{\curly{C}_m}-\frac{1}{m}-\frac{8d}{m^2}\l(1+\frac{n}{m^{d-2}}\r)\r) H_L.
\end{aligned}
$$
Note that we may always lower bound the prefactor on the right-hand side, even if the whole expression is negative, since the resulting inequality is then trivial. Hence, we conclude
$$
H_L^2\geq \frac{m^d}{(m+1)^{d}+n}
\l(\gam_{\curly{C}_m}-\frac{1}{m}-\frac{8d}{m^2}\l(1+\frac{n}{m^{d-2}}\r)\r) H_L.
$$
and now Theorem \ref{thm:fs} follows by frustration-freeness of $H_L$ and the spectral theorem.
\qed

\section{Verification of the finite-size criterion}
\label{sect:verify}
In order to  conclude Theorem \ref{thm:main} from the finite-size criterion in Theorem \ref{thm:fs}, we aim to prove that, for some $m\geq 4$ and a sufficiently small $\de>0$,
\beq\label{eq:threshold}
\gam_{\curly{C}_m}>\frac{1}{m}+\frac{8d}{m^2}\l(1+\frac{n}{m^{d-2}}\r).
\eeq
Such a bound is the content of the following proposition.

\be{prop}
\label{prop:verify}
Define $C_{m,n}:=8((m+1)^d+n)(n^2+n)$ and suppose that $3C_{m,n}\de<1$. Then 
\beq
\gam_{\curly{C}_m}> 1-3C_{m,n}\de.
\eeq
\e{prop}

Notice that the lower bound $1-3C_{m,n}\de$ only exceeds the gap threshold from Theorem \ref{thm:fs} for sufficiently small $\de$. As mentioned before, the proof of Proposition \ref{prop:verify} is based on a perturbative argument via the reference Hamiltonian with $\de=0$. 

\subsection{Proof of the main result assuming Proposition \ref{prop:verify}}
We will now prove Theorem \ref{thm:main} assuming Proposition \ref{prop:verify}. By the finite-size criterion in Theorem \ref{thm:fs}, the claimed gap exists if for some $m\geq 4$ we can verify
$$
\gam_{\curly{C}_m}>\frac{1}{m}+\frac{8d}{m^2}\l(1+\frac{n}{m^{d-2}}\r).
$$
By Proposition \ref{prop:verify}, this is ensured by the condition
\beq\label{eq:deltacondition}
\de<\frac{1-\frac{1}{m}-\frac{8d}{m^2}\l(1+\frac{n}{m^{d-2}}\r)}{3C_{m,n}}.
\eeq
For $d=2$, we can set $m=16n$ to see that \eqref{eq:deltacondition} is guaranteed by the stronger condition
\beq\label{eq:2dchoice}
\de<\frac{1-\frac{3}{16n}}{96n^2(16n+1)^d}
\eeq
so the right-hand side is a possible choice for $c_{2,n}>0$.

 Let $d=3$. When we assume that $m\geq \max\{n^{1/(d-2)},8d\}$, we see that \eqref{eq:deltacondition} is guaranteed by the stronger condition
$$
\de<\frac{1-\frac{3}{m}}{96n^2(m+1)^d}, 
$$
so the right-hand side is a possible choice for $c_{d,n}>0$. This proves Theorem \ref{thm:main}.
\qed

\subsection{The reference Hamiltonian with $\de=0$}
Recall that the original interaction terms $h^{(k)}$ are defined by \eqref{eq:hkdefn} as a sum of projections onto the (normalized) vectors
$$
\phi_i^{(k)}:=\ket{0\otimes i}-\lam_i^{(k)}\ket{i\otimes 0},
\qquad \phi_{ij}^{(k)}:=\lam_i^{(k)}\ket{i\otimes j}-\lam_j^{(k)}\ket{j\otimes i},
\qquad \phi_{ii}^{(k)}:=\ket{i\otimes i}
$$
with $1\leq i< j\leq n$ and $1\leq k\leq d$. Since the vectors are normalized to form $h^{(k)}$, we may, without changing $h^{(k)}$, replace $\phi_{ij}^{(k)}$ by the vector
$$
\varphi_{ij}^{(k)}:=\ket{i\otimes j}-\frac{\lam_j^{(k)}}{\lam_i^{(k)}}\ket{j\otimes i}.
$$
We recall our assumption \eqref{eq:assumedelta}, which implies that, for all $i<j$ and all $1\leq k\leq d$, it holds that
\beq\label{eq:4delta}
\frac{1}{4}\de^{j-i}\leq \frac{\lam_j^{(k)}}{\lam_i^{(k)}}\leq 4\de^{j-i},
\eeq
and this explains how we should define the reference model with $\de=0$. To this end, we first define the reference interaction with $\de=0$ by
\beq\label{eq:tildehkdefn}
\tilde h:=\sum_{i=1}^n \ketbra{\tilde\phi_i}{\tilde\phi_i}
+\sum_{1\leq i\leq j\leq n} \ketbra{\tilde\phi_{ij}}{\tilde\phi_{ij}}
\eeq
where we defined the vectors $\tilde\phi_i,\tilde \phi_{ij}\in \C^{n+1}\otimes \C^{n+1}$ by
\beq\label{eq:tildephidefn}
\tilde\phi_i:=\ket{0\otimes i},
\qquad \tilde\phi_{ij}:=\ket{i\otimes j},
\eeq
for all $1\leq i\leq j\leq n$. We remark that $\tilde h$ is a projection. We generalize Definition \eqref{eq:HSdefn} in the natural way. I.e., given a subgraph $\curly{G}\subset \Lam_L$, we define the associated Hamiltonian
$$
\tilde H_\curly{G}:=\sum_{e\in \curly{E}_{\curly{G}}} \tilde h_e.
$$
We note that $\tilde H_\curly{G}$ is frustration-free (an explicit ground state is given by the tensor product of the state $\ket{0}$ over any subgraph $\curly{G}$). We write $\tilde\gam_{\curly{G}}$ for its spectral gap.

The key fact about the reference Hamiltonian is that it has a large gap of $1$, uniformly in the system size.

\be{lm}[The reference Hamiltonian always has gap $1$]\label{lm:gap}
For all subgraphs $\curly{G}\subset \Lam_L$, it holds that
$$
\tilde \gam_\curly{G}=1.
$$
\e{lm}

\be{proof}
Notice that we may write
$$
\ketbra{\tilde{\phi}_i}{\tilde{\phi}_i}=\ketbra{0}{0}\otimes \ketbra{i}{i},
\qquad \ketbra{\tilde{\phi}_{ij}}{\tilde{\phi}_{ij}}=\ketbra{i}{i}\otimes \ketbra{j}{j}.
$$
This observation implies that all projections $\tilde h_e$ commute, and the lemma follows.
\e{proof}

We recall from the discussion in Section \ref{sect:strategy} that the key fact about the box on a stick is that it does not possess any artificial ground state degeneracies, i.e., no ground state degeneracies are lifted as we pass from $\de=0$ to $\de>0$. We establish this in two steps: In step 1, we characterize all the ground states of the reference Hamiltonian with $\de=0$. In step 2, we use symmetry (particle number conservation) to argue that none of these degeneracies are lifted. Afterwards, we combine these facts with Lemma \ref{lm:gap} to prove Proposition \ref{prop:verify} and hence the main result, Theorem \ref{thm:main}.

\subsection{Step 1: Characterizing the ground states of the reference Hamiltonian}
The following lemma characterizes the ground states of $H_{\curly{C}_m}$. For this, we will use a product basis for the Hilbert space $\curly{H}_{\curly{C}_m}=\bigotimes_{x\in \curly{C}_m} \C^{n+1}$ in the following form
$$
B=\setof{\ket{i_0\otimes i_1 \otimes \ldots  \otimes i_n \otimes \ul{j}}}{i_0,\ldots,i_n\in \{1,\ldots,n\},\ \ul{j}\in \{1,\ldots,n\}^{|\curly{B}_m |-1}},
$$
with the convention that the index $i_l$ describes the state at site $l e_1\in \Lam_L$. In other words, the indices $i_0,\ldots,i_n$ describe the basis states along the stick part of $\curly{C}_m$, while the remaining $\ul{j}$ vector describes the basis state across the box $\curly{B}_m$ without its bottom corner. Notice that the model Hamiltonian $\tilde H_{\curly{C}_m}$ is diagonal in the $B$-basis.

\be{lm}[Ground states of $\tilde H_{\curly{C}_m}$]\label{lm:gs}
Let  $b=\ket{i_0\otimes i_1 \otimes \ldots  \otimes i_n \otimes \ul{j}}\in B$. Then $b\in \ker \tilde H_{\curly{C}_m}$ if and only if $\ul{j}=\ket{0}^{\otimes (|\curly{B}_m |-1)}$ and the sequence $(i_0,i_1,i_2,\ldots,i_n)$ satisfies the following two conditions:
\be{enumerate}[label=(\roman*)]
\item The sequence $(i_0,i_1,i_2,\ldots,i_n)$ is strictly decreasing until it reaches $0$.
\item After the sequence $(i_0,i_1,i_2,\ldots,i_n)$ reaches $0$, it remains at $0$.
\e{enumerate}
\e{lm}

\be{proof}
Fix an edge $e=(le_1,(l+1)e_1)$ with $0\leq l\leq n-1$ along the stick. Notice that $e$ is oriented upwards.  Hence, the projections \eqref{eq:tildephidefn} assign an energy penalty of $1$ at $e$ if and only if $0\neq i_l\leq i_{l+1}$. Conversely, the energy penalty at $e$ is $0$ if and only if $i_l>i_{l+1}$ or $i_l=i_{l+1}=0$. Combining this fact over all  $0\leq l\leq n-1$, we see that $b$ is a ground state across the stick if and only if conditions (i) and (ii) are satisfied. 

Next, observe that conditions (i) and (ii) collectively imply that $i_n=0$. From this we see that the choice $\ul{j}=\ket{0}^{\otimes (|\curly{B}_m |-1)}$ indeed yields a ground state. Conversely, fix an arbitrary index $j_l$ from $\ul{j}$. There is an oriented path in $\curly{B}_m$ that connects the vertex at $ne_1$, which carries the state $\ket{i_n}=\ket{0}$, to the vertex with index $j_l$. Hence, if $j_l\neq 0$, then along that path the Hamiltonian $\tilde H_{\curly{C}_m}$ must incur an energy penalty of at least $1$ from having an oriented edge across which the indices increase. This proves that only the choice $\ul{j}=\ket{0}^{\otimes (|\curly{B}_m |-1)}$ leads to a ground state and finishes the proof of Lemma \ref{lm:gs}.
\e{proof}

\subsection{Step 2: Stability of ground state degeneracies by symmetry}

The crucial observation is now that the ground states of $\tilde H_{\curly{C}_m}$, as characterized in Lemma \ref{lm:gs}, each belong to a unique particle number sector. Since the perturbation of turning on $\de>0$ respects particle number conservation, this means that the ground state degeneracy of $\tilde H_{\curly{C}_m}$ is not lifted.

For each $1\leq i\leq n$, we let $N_i$ be the particle number operator for species $i$, which is defined on a basis element $b=\ket{i_0\otimes i_1 \otimes \ldots  \otimes i_n \otimes \ul{j}}\in B$ as the number of occurrences of the label $i$, i.e., $N_i$ has eigenvalues $\{0,1,\ldots,|\curly{C}_m|\}$. We define the particle number sectors
$$
\curly{P}(\nu_1,\ldots,\nu_n):=\setof{\psi\in \curly{H}_{\curly{C}_m}}{N_i \psi= \nu_i \psi,\,\forall 1\leq i\leq n},
$$
for every sequence of particle numbers $(\nu_1,\ldots,\nu_n)\in \{0,1,\ldots,|\curly{C}_m|\}^n$. Note that both Hamiltonians, $H_{\curly{C}_m}$ and $\tilde H_{\curly{C}_m}$, commute with each particle number operator $N_i$ and are hence block-diagonal with respect to the particle number sectors $\curly{P}(\nu_1,\ldots,\nu_n)$. In fact, it was observed in \cite{BHNY}, by explicitly constructing the ground states of the true Hamiltonian $H_{\curly{C}_m}$ for any choice of anisotropy parameters $\lam_{i}^{(k)}\neq 0$ that there are $2^n$ ground states which each lie in a unique particle number sector $\curly{P}(\nu_1,\ldots,\nu_n)$ with $(\nu_1,\ldots,\nu_n)\in \{0,1\}^n$.  In other words, the ground states of $H_{\curly{C}_m}$ are labeled by subsets $M\subset \{1,\ldots,n\}$ and $M$ encodes the particle species which are (once) present in the ground state.

We now observe that the same fact is true for the reference Hamiltonian $\tilde H_{\curly{C}_m}$.

\be{cor}[of Lemma \ref{lm:gs}]\label{cor:symm}
The ground states of $\ker \tilde H_{\curly{C}_m}$ are labeled by subsets $M\subset \{1,\ldots,n\}$. More precisely, for each ground state $b\in B$ of $\ker \tilde H_{\curly{C}_m}$, there exists a unique vector $(\nu_1,\ldots,\nu_n)\in \{0,1\}^n$ such that $b \in \curly{P}(\nu_1,\ldots,\nu_n)$.
\e{cor}

\be{proof}
By Lemma \ref{lm:gs}, $\ul{j}=\ket{0}^{\otimes (|\curly{B}_m |-1)}$ for ground states and so particles (i.e. non-zero indices) can only occur along the stick. The strictly decreasing condition (ii) from Lemma \ref{lm:gs} then implies the corollary.
\e{proof}

Corollary \ref{cor:symm} allows us to apply non-degenerate perturbation theory in each particle number sector and conclude the proof of Proposition \ref{prop:verify}.

\subsection{Conclusion}
We will need the following lemma on the size of the perturbation in passing from the reference Hamiltonian $\tilde H_{\curly{C}_m}$ to the true Hamiltonian $H_{\curly{C}_m}$.

\be{lm}\label{lm:norm}
We have the norm bound
$$
\|H_{\curly{C}_m}-\tilde H_{\curly{C}_m}\|\leq C_{m,n}  \de,\qquad C_{m,n}:=8((m+1)^d+n)(n^2+n).
$$
Consequently, for every normalized $\phi\in\curly{H}_{\curly{C}_m}$, we have
\beq\label{eq:conseq}
\scp{\phi}{H_{\curly{C}_m}\phi}\geq \scp{\phi}{\tilde H_{\curly{C}_m}\phi}-C_{m,n}\de.
\eeq
\e{lm}

\be{proof}
We first note that, for any edge $e_0\in \curly{E}_{\curly{C}_m}$,
$$
\begin{aligned}
&\|H_{\curly{C}_m}-\tilde H_{\curly{C}_m}\|
=\l\|\sum_{e\in \curly{E}_{\curly{C}_m}} (h_e-\tilde h_e)\r\|\\
&\leq |\curly{E}_{\curly{C}_m}| \|h_{e_0}-\tilde h_{e_0}\|
=((m+1)^d+n) \|h_{e_0}-\tilde h_{e_0}\|.
\end{aligned}
$$
Notice that $h_{e_0}-\tilde h_{e_0}$ consists of a total of $\binom{n}{2}+n=\frac{n^2+n}{2}$ differences between projections. These are differences between projections onto $\ket{\hat{\phi}_{i,j}^{(k)}}$, respectively $\ket{\tilde\phi_{i,j}}$, with $i\neq j$, and differences between projections onto $\ket{\hat{\phi}_{i}^{(k)}}$, respectively $\ket{\tilde \phi_{i}}$. Thanks to \eqref{eq:4delta} and the general estimate $\|\ketbra{v}{v}-\ketbra{w}{w}\|\leq 2\|\ket{v}-\ket{w}\|$ for normalized vectors $\|v\|=\|w\|=1$, the norm of these differences is bounded by 
$$
\frac{16\de}{\sqrt{1+\l(\frac{\de}{4}\r)^2}}\leq 16\de.
$$
This shows
$$
\|h_{e_0}-\tilde h_{e_0}\|\leq 8\de (n^2+n)
$$
and proves the norm bound. Finally, we obtain \eqref{eq:conseq} by
$$
\begin{aligned}
\scp{\phi}{H_{\curly{C}_m}\phi}
=\scp{\phi}{\tilde H_{\curly{C}_m}\phi}+\scp{\phi}{(H_{\curly{C}_m}-\tilde H_{\curly{C}_m})\phi}
\geq \scp{\phi}{\tilde H_{\curly{C}_m}\phi}-C_{m,n}  \de
\end{aligned}
$$
and Lemma \ref{lm:norm} is proved.
\e{proof}

We are now ready to conclude the argument, using non-degenerate perturbation theory on a fixed particle number subspace (case 2 below).

\be{proof}[Proof of Proposition \ref{prop:verify}]
Let $\psi\in (\ker H_{\curly{C}_m})^\perp$ be a normalized eigenstate of $H_{\curly{C}_m}$. Without loss of generality, we may also assume that $\psi \in \curly{P}(\nu_1,\ldots,\nu_n)$ for some choice of $(\nu_1,\ldots,\nu_n)\in \{0,1,\ldots,|\curly{C}_m|\}^n$. 

\textbf{Case 1:} Let $\psi \in \curly{P}(\nu_1,\ldots,\nu_n)$ such that at least one $\nu_l>1$. By Corollary \ref{cor:symm}, this implies that $\psi \in (\ker \tilde H_{\curly{C}_m})^\perp$. Hence, by Lemmas \ref{lm:norm} and \ref{lm:gap},
$$
\begin{aligned}
\scp{\psi}{H_{\curly{C}_m}\psi}
\geq\scp{\psi}{\tilde H_{\curly{C}_m}\psi}-C_{m,n}  \de
\geq 1-C_{m,n}  \de
\end{aligned}
$$
This concludes case 1.

\textbf{Case 2:} Let $\psi \in \curly{P}(\nu_1,\ldots,\nu_n)$ with each $\nu_l\in \{0,1\}$. Let us write $\psi_0$, respectively $\tilde\psi_0$, for the unique ground state of $\ker H_{\curly{C}_m}$, respectively $ \ker \tilde H_{\curly{C}_m}$,  in the sector $\curly{P}(\nu_1,\ldots,\nu_n)$ that has non-negative components in the tensor product basis $B$. (Note that this used Corollary \ref{cor:symm} and the fact that the explicit ground states from \cite{BHNY} also have non-negative components.) Then we decompose $\psi$ as follows:
$$
\psi=c_0\tilde\psi_0+\psi_\perp,\qquad c_0:=\scp{\psi}{\tilde\psi_0},
$$
with $\psi_\perp\in\curly{P}(\nu_1,\ldots,\nu_n)\cap (\ker \tilde  H_{\curly{C}_m})^\perp$. By Lemma \ref{lm:norm}, $\tilde  H_{\curly{C}_m}\tilde\psi_0=0$, and Lemma \ref{lm:gap}, we find
\beq\label{eq:sim}
\begin{aligned}
\scp{\psi}{H_{\curly{C}_m}\psi}
\geq& \scp{\psi}{\tilde H_{\curly{C}_m}\psi}-C_{m,n}\de\\
=&\scp{\psi_\perp}{\tilde H_{\curly{C}_m}\psi_\perp}-C_{m,n}\de\\
\geq&\|\psi_\perp\|^2-C_{m,n}\de\\ 
=&1-|c_0|^2-C_{m,n}\de.
\end{aligned}
\eeq
It remains to bound $|c_0|$. We recall that $\psi\in (\ker H_{\curly{C}_m})^\perp$ and so $\scp{\psi}{\psi_0}=0$. By Cauchy-Schwarz,
$$
|c_0|^2
=|\scp{\psi}{\tilde\psi_0}|^2
=|\scp{\psi}{\tilde\psi_0-\psi_0}|^2
\leq \|\tilde\psi_0-\psi_0\|^2.
$$
It remains to bound $\|\tilde\psi_0-\psi_0\|^2$. The following lemma rests on the uniqueness of ground states in the sector $\curly{P}(\nu_1,\ldots,\nu_n)$, and hence on Corollary \ref{cor:symm}.

\be{lm}\label{lm:degdiff}
It holds that $\|\tilde\psi_0-\psi_0\|^2\leq 2C_{m,n}\de$.
\e{lm}

We will prove this lemma below. For now, we note that it implies $|c_0|^2\leq 2C_{m,n}\de$ and so, by \eqref{eq:sim}, also
$$
\scp{\psi}{H_{\curly{C}_m}\psi}\geq 1-3C_{m,n}\de,
$$
which verifies the claim of Proposition \ref{prop:verify} also in case 2. It thus remains to prove the lemma.\\

\noindent\textit{Proof of Lemma \ref{lm:degdiff}.}
We write 
$$
\psi_0=a_0\tilde\psi_0+\varphi_\perp,\qquad a_0:=\scp{\psi_0}{\tilde\psi_0},
$$
with $\varphi_\perp\in\curly{P}(\nu_1,\ldots,\nu_n)\cap (\ker \tilde  H_{\curly{C}_m})^\perp$. Notice that $a_0=\scp{\psi_0}{\tilde \psi_0}\geq 0$, since $\psi_0$ and $\tilde\psi_0$ have non-negative components in the $B$-basis. On the one hand, we can argue similarly as in \eqref{eq:sim} by using Lemma \ref{lm:norm}, $\tilde  H_{\curly{C}_m}\tilde\psi_0=0$, and Lemma \ref{lm:gap} to find
\beq\label{eq:sim'}
\begin{aligned}
0=\scp{\psi_0}{H_{\curly{C}_m}\psi_0}
\geq \scp{\varphi_\perp}{\tilde H_{\curly{C}_m}\varphi_\perp}-C_{m,n}\de
\geq 1-a_0^2-C_{m,n}\de,
\end{aligned}
\eeq
which is equivalent to $a_0\geq \sqrt{1-C_{m,n}\de}\geq 1-C_{m,n}\de$. On the other hand, using that $a_0\geq0$, we have
$$
\|\tilde\psi_0-\psi_0\|^2=2-2\Re \scp{\psi_0}{\tilde \psi_0}= 2-2a_0\leq 2C_{m,n}\de.
$$
This proves Lemma \ref{lm:degdiff}, and hence Proposition \ref{prop:verify}.
\e{proof}

\section*{Acknowledgements}

The authors thank Amanda Young for helpful comments. B.N.\ acknowledges support by the National Science Foundation under Grant DMS-1813149 and a CRM-Simons
Professorship for a stay at the Centre de Recherches Math\'ematiques (Montr\'eal) during Fall 2018 where part of this work was carried out.

\end{document}